\documentstyle[11pt,epsf,twocolumn]{article}

\newcommand{\miset} 	{\mbox{$\not\!\!{E_T}$}}
\newcommand{\sigmiset}  {S(\miset)}
\newcommand{\twb } 	{t{\rightarrow}W^{+}b}
\newcommand{\thb } 	{t{\rightarrow}H^{+}b}
\newcommand{\tthh} 	{t\bar{t}{\rightarrow}H^+H^-b\bar{b}}
\newcommand{\tthw} 	{t\bar{t}{\rightarrow}H^{\pm}W^{\mp}b\bar{b}}
\newcommand{\ttww} 	{t\bar{t}{\rightarrow}W^+W^-b\bar{b}}                          
\newcommand{\et  } 	{E_T}
\newcommand{\mt  } 	{m_{\rm top}}

\begin{document}

\onecolumn

\begin{flushright}
FERMILAB-PUB-96/004-E \\
(Submitted to Phys. Rev. D) \\
8 January 1995 \\
\end{flushright}

\vskip 1.0cm
\begin{center}
{\LARGE \bf {Search for charged Higgs decays of the top quark
            using hadronic tau decays}}
\end{center}

\font\eightit=cmti8
\def\r#1{\ignorespaces $^{#1}$}
\hfilneg
\begin{sloppypar}
\noindent
F.~Abe,\r {14} H.~Akimoto,\r {32}
A.~Akopian,\r {27} M.~G.~Albrow,\r 7 S.~R.~Amendolia,\r {23} 
D.~Amidei,\r {17} J.~Antos,\r {29} C.~Anway-Wiese,\r 4 S.~Aota,\r {32}
G.~Apollinari,\r {27}
% ...
\end{sloppypar}

\vskip .025in
\begin{center}
(CDF Collaboration)
\end{center}

\vskip .025in
\begin{center}
\r 1  {\eightit Argonne National Laboratory, Argonne, Illinois 60439} \\
\r 2  {\eightit Istituto Nazionale di Fisica Nucleare, University of Bologna,
I-40126 Bologna, Italy} \\
\r 3  {\eightit Brandeis University, Waltham, Massachusetts 02254} \\
\r 4  {\eightit University of California at Los Angeles, Los 
Angeles, California  90024} \\  
\r 5  {\eightit University of Chicago, Chicago, Illinois 60637} \\
\r 6  {\eightit Duke University, Durham, North Carolina  27708} \\
\r 7  {\eightit Fermi National Accelerator Laboratory, Batavia, Illinois 
60510} \\
\r 8  {\eightit Laboratori Nazionali di Frascati, Istituto Nazionale di Fisica
               Nucleare, I-00044 Frascati, Italy} \\
\r 9  {\eightit Harvard University, Cambridge, Massachusetts 02138} \\
\r {10} {\eightit Hiroshima University, Higashi-Hiroshima 724, Japan} \\
\r {11} {\eightit University of Illinois, Urbana, Illinois 61801} \\
\r {12} {\eightit Institute of Particle Physics, McGill University, Montreal 
H3A 2T8, and University of Toronto,\\ Toronto M5S 1A7, Canada} \\
\r {13} {\eightit The Johns Hopkins University, Baltimore, Maryland 21218} \\
\r {14} {\eightit National Laboratory for High Energy Physics (KEK), Tsukuba, 
Ibaraki 305, Japan} \\
\r {15} {\eightit Lawrence Berkeley Laboratory, Berkeley, California 94720} \\
\r {16} {\eightit Massachusetts Institute of Technology, Cambridge,
Massachusetts  02139} \\   
\r {17} {\eightit University of Michigan, Ann Arbor, Michigan 48109} \\
\r {18} {\eightit Michigan State University, East Lansing, Michigan  48824} \\
\r {19} {\eightit University of New Mexico, Albuquerque, New Mexico 87131} \\
\r {20} {\eightit Osaka City University, Osaka 588, Japan} \\
\r {21} {\eightit Universita di Padova, Istituto Nazionale di Fisica 
          Nucleare, Sezione di Padova, I-35131 Padova, Italy} \\
\r {22} {\eightit University of Pennsylvania, Philadelphia, 
        Pennsylvania 19104} \\   
\r {23} {\eightit Istituto Nazionale di Fisica Nucleare, University and Scuola
               Normale Superiore of Pisa, I-56100 Pisa, Italy} \\
\r {24} {\eightit University of Pittsburgh, Pittsburgh, Pennsylvania 15260} \\
\r {25} {\eightit Purdue University, West Lafayette, Indiana 47907} \\
\r {26} {\eightit University of Rochester, Rochester, New York 14627} \\
\end{center}

\twocolumn[
\begin{center}
\r {27} {\eightit Rockefeller University, New York, New York 10021} \\
\r {28} {\eightit Rutgers University, Piscataway, New Jersey 08854} \\
\r {29} {\eightit Academia Sinica, Taipei, Taiwan 11529, Republic of China} \\
\r {30} {\eightit Texas A\&M University, College Station, Texas 77843} \\
\r {31} {\eightit Texas Tech University, Lubbock, Texas 79409} \\
\r {32} {\eightit University of Tsukuba, Tsukuba, Ibaraki 305, Japan} \\
\r {33} {\eightit Tufts University, Medford, Massachusetts 02155} \\
\r {34} {\eightit University of Wisconsin, Madison, Wisconsin 53706} \\
\r {35} {\eightit Yale University, New Haven, Connecticut 06511} \\
\end{center}

\vskip 1.5cm
\begin{center} 
\small{Abstract}
\end{center}

\small{
We present the result of a search for charged Higgs decays of the top
quark, produced in $p\bar{p}$ collisions at $\surd s =$
1.8~TeV. When the charged Higgs is heavy and decays to a tau lepton,
which subsequently decays hadronically, the resulting events have a
unique signature: large missing transverse energy and the
low-charged-multiplicity tau.  Data collected in the period 1992-1993
at the Collider Detector at Fermilab, corresponding to
18.7$\pm$0.7~pb$^{-1}$, exclude new regions of combined top quark and
charged Higgs mass, in extensions to the standard model with two
Higgs doublets.
}

\vskip0.5in
]

\begin{center}
\section{\normalsize INTRODUCTION}
\end{center}

We have conducted a search for decays of the top quark to a charged
Higgs boson, using the Higgs decays to hadronically decaying tau
leptons.  The results presented here come from data collected during
the years 1992--1993 at the Collider Detector at Fermilab,
corresponding to an integrated luminosity of $18.7\pm0.7$ pb$^{-1}$. A
charged Higgs arises in extensions to the Standard Model with two
Higgs doublets~\cite{two_higgs}.  If the charged Higgs exists in such
a model and is lighter than the top quark, then two competing channels
are possible: $\thb$ and $\twb$. The charged Higgs can decay either to
$\tau\bar{\nu}$ or to $c\bar{s}$. The branching ratios of these
processes depend on the top quark and charged Higgs masses, and on
$\tan\beta$, the ratio of the vacuum expectation values of the two
Higgs doublets in the model.\footnote{We consider here only models in
which one Higgs doublet couples to the up-type quarks, and the other
doublet couples to the down-type quarks and the leptons.} We consider
here only the kinematically allowed cases where $\mt > m_H + m_b$ and
$\mt > m_W + m_b$. In these cases three decay modes of the quark pairs
are possible: $\tthh$, $\tthw$, and $\ttww$.

\begin{figure}[t]
  \parbox{3.25in}{\epsfxsize=3.25in\epsffile[60 0 320 240]{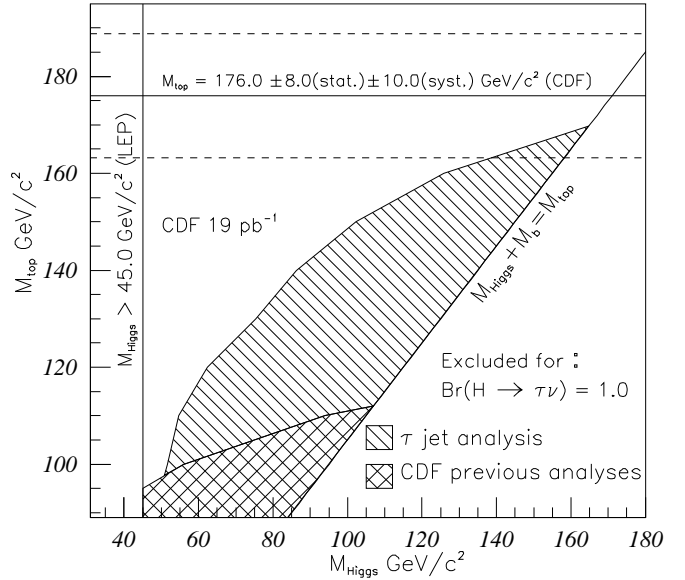}}
  \caption{Regions of the ($m_{\rm top}$,$m_{\rm Higgs}$) plane
           excluded at 95\% CL for a 100\% branching ratio of 
           $H^\pm\rightarrow\tau\nu$. The plot
           also shows the limit from the previous 
           analyses~{\protect \cite{CDF2164,CDF2435}}.}
  \label{masslarge}
\end{figure}

 This analysis of hadronic decays of the $\tau$-lepton uses a method
similar to that used previously~\cite{CDF2164} but with
four times more integrated luminosity and an event selection designed
for larger top and charged Higgs masses.  Our most recent
limit~\cite{CDF2435} uses the leptonic decays of the
tau using the data set collected in the same period as in this paper,
i.e.  1992--1993.  Charged Higgs masses from 45 GeV/$c^2$ to 110
GeV/$c^2$, and top quark masses from 90 GeV/$c^2$ to 110
GeV/$c^2$ were excluded at a 95\% CL, as shown in the lower hatched
part of the plot in Fig.~\ref{masslarge}.  Experiments at LEP exclude
a charged Higgs with mass less than 45 GeV/$c^2$~\cite{limit}.

In the present analysis a more stringent limit results from the 64\%
hadronic branching ratio of the tau, compared with the 36\% leptonic
branching ratio.  However, the larger expected background, mainly
hadronic processes, must be well modelled.  The analysis presented
here addresses top masses in the range extending from the limits of
previous searches~\cite{CDF2164,CDF2435}, about 100 GeV/$c^2$, up to
the mass range which has been measured, 176$\pm$8(stat)$\pm$10(syst)
GeV/$c^2$~\cite{CDF_top}.  The analysis excludes by direct search a
top or top-like object decaying via a charged Higgs in this region.

Top quark pair events with one or two charged Higgs decays should
contain energetic jets which come from $b$ quarks and the decays of
the taus.  Each top quark leads to the production of two energetic
neutrinos, leading in turn to a large missing transverse energy,
denoted $\miset$.\footnote{For a calorimeter energy deposit, assuming
the particles came from some point along the beam axis, a direction in
space is defined.  The transverse energy $\et$ is the component of the
energy vector in the plane perpendicular to the beam axis.  The
$\miset$ is defined as the magnitude of the vector sum of the
transverse energy $\et$ of each calorimeter energy deposit.} The first
neutrino is emitted in the charged Higgs or W boson decay, and the
second results from the tau decay.  The signature of an hadronically
decaying tau is a narrow jet with either one or three associated
charged particles.  Thus, top quark pair events with charged Higgs
decays to tau leptons can be found by looking for an excess of events
with narrow energetic jets with one or three charged particles coming
from the tau, along with the presence of other energetic hadronic jets
and neutrinos.  This signature differs from that of standard model top
events due to the higher probability of decays to tau leptons, and
larger missing $\et$.

\begin{center}
\section{\normalsize{CDF DETECTOR AND TRIGGER}}
\end{center}

The CDF detector is described in detail
elsewhere~\cite{CDF_description}. The most important components of the
CDF detector for this analysis are the tracking chambers and
calorimeters.  The relevant tracking chambers are the vertex time
projection chamber (VTX) and the central tracking chamber (CTC), which
is a large cylindrical drift chamber surrounding the VTX. Both are
located inside a superconducting solenoid magnet generating a 1.4~T
field.  The VTX provides z-vertex reconstruction and $r-z$ tracking
over the pseudorapidity range $|\eta| < 3.25$~\cite{define_eta}, where
the $z$ axis is the proton direction along the beam line and $r$
refers to the radial coordinate transverse to the beam line. The
momenta of charged particles are measured in the CTC. The solenoid and
the tracking volume of CDF lie inside electromagnetic and hadronic
calorimeters which cover $2\pi$ in azimuth and up to $|\eta|= 4.2$.
The calorimeters are segmented in azimuth and pseudorapidity to form a
tower geometry which points back to the nominal interaction point $z=0$.

The ``trigger'' decision as to whether or not the data from a
particular interaction should be recorded depends on the particular
pattern of energy deposited in the calorimeters, the presence of
charged tracks in the CTC, and the presence of penetrating charged
particles in the muon chambers which surround the calorimeter.  This
analysis relies in particular on a trigger which uses analog sums of
the calorimeter energy deposits to determine the missing transverse
energy.  Since the charged Higgs events sought in this analysis
generally have large missing transverse energy, this analysis uses
only those events which satisfy a trigger requirement of at least
35~GeV of missing transverse energy.

\begin{center}
\section{\normalsize{EVENT SELECTION}}
\end{center}

The criteria to reject background and to select the charged Higgs
signal were determined using a Monte Carlo simulation based on top
quark and charged Higgs masses just beyond those excluded in previous
analyses, namely $m_{\rm top} = 120~$GeV/$c^2$, and $m_{\rm Higgs} =
100~$GeV/$c^2$.  A version of the Monte Carlo program
ISAJET~\cite{isajet}, modified to correctly model the polarization of
the taus, generated events which were then passed through the CDF
detector simulation.

The selection criteria aim to select events with large missing
transverse energy due to the neutrinos, the presence of a hadronically
decaying tau lepton, and at least one other jet due either to another
tau or to one of the jets from the top quark.  Each event must have

\begin{itemize}
  \item $\miset >$ 40 GeV,
  \item $\sigmiset >$ 4 GeV$^{1/2}$, 
  \item a tau lepton identified as discussed below, with 
        \begin{itemize}
          \item $\et >$ 30 GeV,
          \item $|\eta| < 1$,
        \end{itemize}
  \item a jet as defined below, with 
        \begin{itemize}
          \item $\et >$ 20 GeV,
          \item $|\eta| < 2$,
        \end{itemize}
  \item $\Delta\phi_{\tau-jet} < 140^\circ$, and
  \item scalar $\Sigma|E_{T}| > 100$ GeV,
\end{itemize}
where we use the definition $\sigmiset\equiv\miset/\sqrt{\Sigma|E_T|}$
for the ``significance'' of the missing $\et$.

\begin{figure}[t]
  \parbox{3.25in}{\epsfxsize=3.25in\epsffile[60 0 320 240]{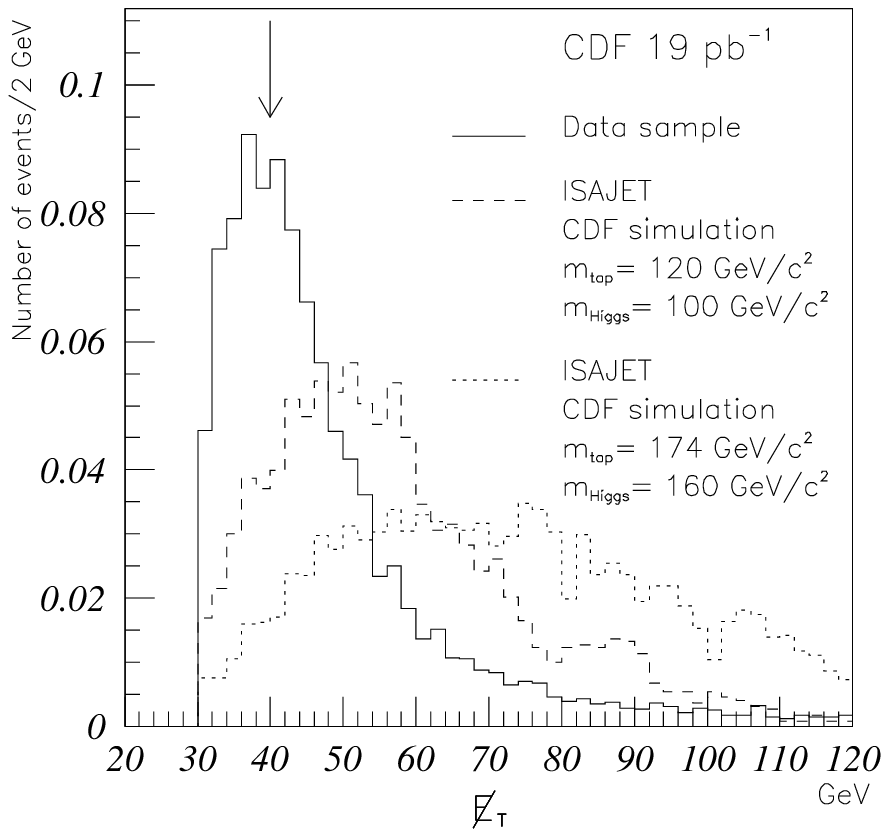}}
  \caption{Distribution of $\miset$
           for the data sample (solid line) and for the Monte Carlo
           simulation (dashed and dotted line). Events to the left of the arrow
           are removed by the cuts on the data sample.}
  \label{missing_energy}
\end{figure}

\begin{figure}[t]
  \parbox{3.25in}{\epsfxsize=3.25in\epsffile[60 0 320 240]{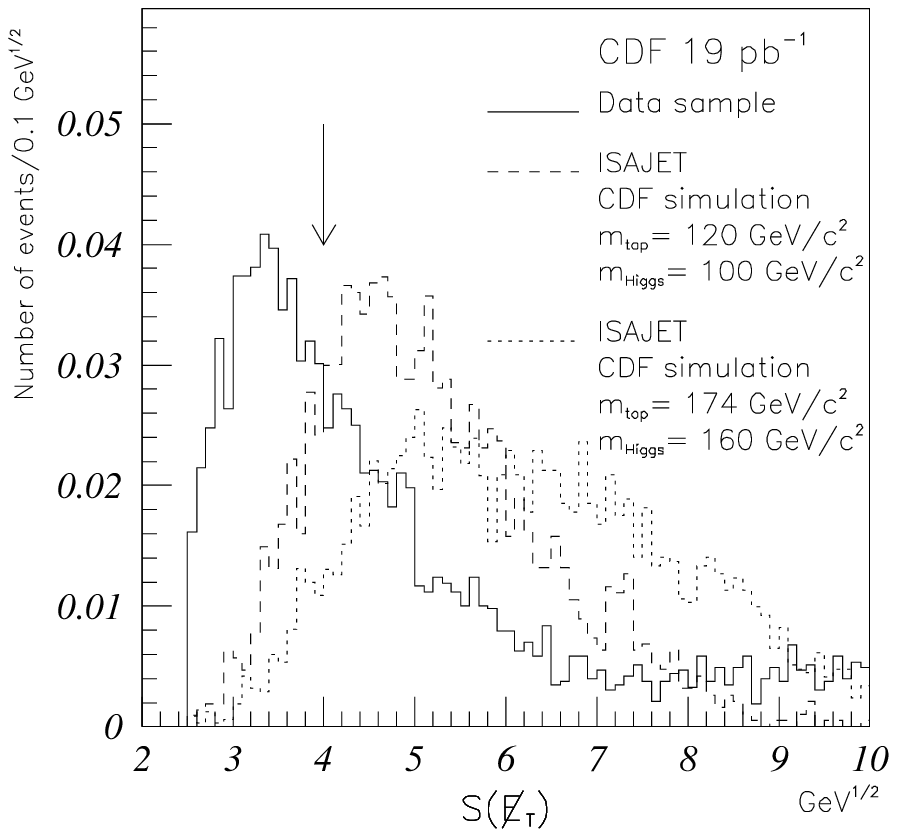}}
  \caption{Distribution of $\sigmiset$ for the data sample (solid line) 
           and for the Monte Carlo simulations (dashed and dotted line).
           Events to the left of the arrow are removed by the cut.}
  \label{missing_energy_sig}
\end{figure}

The number of events which satisfy each criterion are listed in Table~I.
The relative efficiencies between consecutive cuts for the
ISAJET Monte Carlo simulation with $m_{\rm top}=120$ GeV/$c^2$, and
$m_{\rm Higgs}=100$ GeV/$c^2$ are also shown in Table~I.

The scalar nature of the charged Higgs implies that the two neutrinos
produced in the decay chain tend to travel in the same direction,
resulting in a large $\miset$.  Furthermore, the charged Higgs decays
mainly to a tau for large values of the parameter $\tan\beta$.  For
smaller $\tan\beta$ values, the probability for the top quark to decay
to a W boson increases, and the charged Higgs decays more often to a
quark-antiquark pair.  In this case the average $\miset$ consequently
becomes smaller.  Thus, the trigger and selection requirements on
$\miset$ enhance the acceptance in the case of large values of
$\tan\beta$.

The criteria on the missing transverse energy and significance
reinforce the trigger requirements and select events with energetic
neutrinos.  The distributions of $\miset$,and $\sigmiset$ are shown in
Fig.~\ref{missing_energy} and Fig.~\ref{missing_energy_sig}, for the
data sample and for the Monte Carlo simulation of the signal.

When a top quark decays to a charged Higgs, a large fraction of its
energy goes into creation of the charged Higgs. The smaller remaining
energy for the $b$ quark produces jets of lower $\et$. Since the
charged Higgs carries a large energy, its decay products receive a
strong boost. In particular, the taus which come from the charged
Higgs have very large $\et$, resulting in a large-transverse-momentum
($p_T$) associated charged particle near the jet axis.  The average
$p_T$ increases with increasing charged Higgs mass.  A charged
particle is associated with a jet if its initial direction points
within a cone of radius $\Delta
R=\sqrt{\Delta\eta^{2}+\Delta\phi^{2}}=0.4$ of the jet direction,
where $\phi$ is the azimuthal angle in the plane perpendicular to the
beam axis.

The event selection requires the presence of at least two jets, formed
from calorimeter energy deposits in a cone of radius $\Delta R = 0.4$.
The first jet must have $\et > 30$~GeV, lie in the region $|\eta|<1$,
and have an associated charged particle with $p_T > 4$ GeV/c.  If more
than one jet satisfies these criteria, the jet with the largest $\et$
is chosen.

The requirements on the second jet are less stringent.  The Monte
Carlo simulation shows that the second jet has a smaller $\et$, and is
less often in the central region of the detector, $|\eta|<1$.  The
second jet must have $\et > 20$ GeV, $|\eta| < 2$, and an associated
charged particle.

Requiring that the $z$-intercept of the largest-$p_T$ charged particle
associated with each jet be within 5~cm of the primary z-vertex of
the event rejects jets from additional interactions in the event.

\begin{table*}[t]

  \begin{center}
  \begin{tabular}{lcr} \hline \hline
Cuts                          &  Remaining Events & Relative Efficiency \\ \hline 
Initial selection             &        7109       &  22.2$\pm$0.4\% \\
$\miset> 40$ GeV              &        4766       &  85.8$\pm$0.8\% \\
Jet 1 $\et > 30$ GeV          &        2579       &  83.5$\pm$0.8\% \\
Jet 2 $\et > 20$ GeV          &        1601       &  81.6$\pm$1.0\% \\
Azimuthal angle between jets  &        1579       & 100.0$^{+0.0}_{-0.1}$\% \\
$\sigmiset > 4$ GeV$^{1/2}$   &         659       &  79.7$\pm$1.1\% \\
Isolation                     &         193       &  48.9$\pm$1.6\% \\
Electron/jet rejection        &         104       &  93.4$\pm$1.1\% \\
$\Delta$z vertex              &          81       & 100.0$^{+0.0}_{-0.2}$\% \\
$\Sigma|E_{T}| > 100$ GeV     &          74       &  93.3$\pm$1.2\% \\ 
Charged multiplicity          &          19       &  88.6$\pm$1.5\% \\ \hline
Total efficiency              &           -       &   3.9$\pm$0.2\% \\ \hline \hline 
  \end{tabular} \\
  \caption{Number of events selected and signal efficiency for each 
           selection criterion.  The efficiencies represent the 
           successive effect of each criterion, for events in a Monte
           Carlo simulation with $m_{\rm top}~=120$ GeV/$c^2$ and
           $m_{\rm Higgs}=100$ GeV/$c^2$.}
  \label{cut_table}
  \end{center}

\end{table*}

Subsequent criteria to identify hadronically decaying taus assign one
of the two jets as the ``tau.''  Since the charged particles in
hadronically decaying energetic taus must lie in a narrow cone around
the calorimeter energy deposit, the tau must satisfy an isolation
criterion in which there must be no associated charged particle with
$p_T >1$ GeV/c found between cones of $10^{\circ}$ and $30^{\circ}$
defined around the direction of the associated track with the largest
$p_T$.  If the first jet fails the cut, the algorithm applies the cut
to the second jet which in addition must then pass the stricter jet
$\et$, associated charged particle $p_T$, and $\eta$ cuts of the first
jet. The jet which passes the isolation cut is called the tau
candidate.

A small fraction of electrons and single hadrons or low-multiplicity
hadronic jets also satisfy the tau selection criteria. To reject
electrons, the tau candidate must satisfy $1-(10E_{T}/\Sigma
|p_{T}|-1)^{-1}>f_{EM}$, where $f_{EM}$ is the fraction of the total
energy deposited in the electromagnetic calorimeter and
$\Sigma|p_{T}|$ is the sum of the magnitudes of the transverse momenta
of charged particles in the $10^\circ$ cone around the jet axis. A
fraction of the hadronic background is rejected by a similar cut:
$1-(b E_{T}/\Sigma p_{T}-1)^{-1}<f_{EM}$, where the factor $b$ has
been optimized as a function of the $\et$ of the tau candidate: $b =
0.815$ for 30 GeV$ \leq E_{T}<45$ GeV, $b = 0.995$ for 45 GeV$ \leq
E_{T}<69$ GeV, and $b = 0.860$ for $E_{T} \geq $ 69 GeV.

Lastly, the requirement that the scalar sum of transverse energy 
exceeds 100 GeV removes background from $W+{\rm jets}$ events in
which the W decays leptonically.

\begin{center}
\section{\normalsize{TAU SIGNAL IN HADRONIC EVENTS}}
\end{center}
  
\begin{figure}[t]
  \parbox{3.25in}{\epsfxsize=3.25in\epsffile[60 0 320 240]{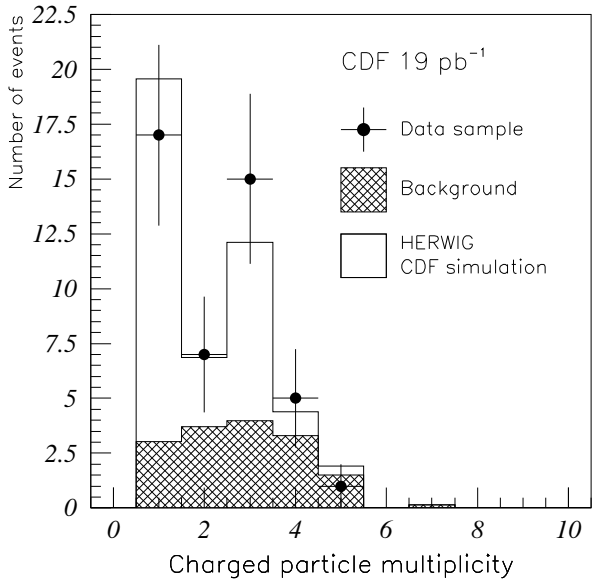}}
  \caption{The tau signal found in the data sample using a similar 
           algorithm with less stringent cuts. The data
           (points) are compared with an estimate of jet background
           (cross-hatched), and with a Monte Carlo simulation of
           W+jets events, with $W \rightarrow \tau\nu$ (open histogram).}
  \label{tau_sig}
\end{figure}

In order to demonstrate that the criteria select hadronic tau decays in a
process known to contain tau leptons,  one can extract the tau signal from 
the process $p\bar{p} \rightarrow W + {\rm jets}, W\rightarrow
\tau\nu$ by selecting a sample of events with a tau, a jet and missing transverse
energy.  Employing less stringent cuts on the $\miset$, $\sigmiset$,
$\et$ and $p_T$ of the charged particles associated with the jets, and
then requiring $\Sigma|E_{T}| < 85$~GeV and a tight cut on the width
of the energy deposit if the jets, results in the multiplicity
distribution shown in Fig.~\ref{tau_sig}.  The plot shows a clear
excess of one and three charged particles in the distribution of the
associated charged particle multiplicity distribution in the
$10^{\circ}$ cone, attributed to W plus jets events.  The data
(points) are compared to a background estimate (cross-hatched
histogram), and a HERWIG Monte Carlo simulation~\cite{herwig} of
$W\rightarrow\tau\nu$ (open histogram), normalized according to the
HERWIG predicted cross section.

\begin{center}
\section{\normalsize{MONTE CARLO SIMULATION AND TRIGGER EFFICIENCY}}
\end{center}

The simulation of charged Higgs events uses the physics generator
ISAJET and the CDF detector simulation.  In order to compute the
acceptance at any point in the top quark versus charged Higgs mass
plane, it is necessary to make a mixture of events from the simulation
of $\tthh$, $\tthw$, and $\ttww$ processes for charged Higgs masses in
the range of 50 GeV/$c^2$--160 GeV/$c^2$ and for top quark masses in
the range of 100 GeV/$c^2$--174 GeV/$c^2$. The $t\bar{t}$ production
cross-section is taken from a next-to-leading order theoretical
calculation~\cite{top_cross}.  The simulation of the effect of the
$\miset$ trigger efficiency comes from a measurement of the efficiency
as a function of the $\miset$ in events which triggered on the
presence of jets.

In the Monte Carlo simulation, more than 95\% of tau candidates found
by the algorithm correspond in spatial direction to the actual taus
generated in the event.

\typeout{Start background section}

\begin{center}
\section{\normalsize{BACKGROUNDS}}
\end{center}

The dominant backgrounds to charged Higgs events in the selected event
sample are hadronic processes, and processes in which a Z or W is
produced, possibly accompanied by jets.  In almost all of the
background, an hadronic jet fluctuates to have low charged particle
multiplicity and satisfies the tau criteria.  A small additional
contribution to the background comes from W and Z events where the tau
jet comes from a tau or mismeasured electron from the boson decay.  In
this case the tau jet typically has one or three associated charged
tracks.

A combination of events satisfying the various jet energy triggers
in the experiment models the hadronic background well.  The background
normalization is computed as a function of the $\et$ and charged
multiplicity of the tau.  The normalization equalizes the number of
events of any charged multiplicity except 1 or 3, in three ranges of
$\et$.  The Monte Carlo simulation shows that real taus contribute
less than a few percent to these bins in multiplicity.  An estimated
total of 17.4$\pm$2.5 events come from processes where the tau jet
came from an hadronic jet; the error is statistical only.

The estimate of the non-hadronic-jet contribution to the background
comes from Monte Carlo simulation of the various processes.  Of these,
only the contribution from $Z\rightarrow\tau^+\tau^-$ remains
non-negligible after all cuts.  Using a total of $30,000$ events
generated with the ISAJET program and then passed through the CDF detector
simulation and analysis, we expect 1.1$\pm$0.4~(stat) events with 1 or
3 associated charged particles.  The production cross section comes
from the measured Z cross section, assuming lepton universality:
$\sigma(p\bar{p}\rightarrow
ZX;Z{\rightarrow}e^{+}e^{-}X) = $0.209 $\pm$ 0.013(stat) $\pm$ 0.017(syst)~nb~
\cite{cross_section}.
This background is small for several reasons: the process has a small
cross-section, the two outgoing taus are azimuthally back-to-back, and
the $\miset$ is typically not large since the tau decay neutrinos are
back-to-back.

The predicted background from W+jets events, with
$W\rightarrow\tau\nu$, comes from 40,000 HERWIG events which were
passed through the CDF detector simulation, including the relevant
trigger efficiencies.  Again assuming lepton universality the
production cross section
$\sigma(p\bar{p}{\rightarrow}WX;W{\rightarrow}e\nu)=2.19 \pm$
0.04(stat)$\pm$0.21 (syst) nb~\cite{cross_section} is used for
normalization.  Most of these events are rejected by the cuts on
$\miset$, $\sigmiset$, and $\Sigma|\et|$.  No event passed the
selection criteria.

The other processes involving W and Z bosons result in background
taken into account by the hadronic jet sample, and contribute 
negligibly to the non-hadronic-jet component.  

There is a small acceptance for events from standard model top quark
pair production; for a top quark with mass of 176 GeV/$c^2$ one
expects 0.2$\pm$0.1 (stat) events.  This acceptance affects the number
of expected events in the signal, and does not enter the background
estimate.

\typeout{Start systematic section}

\begin{center}
\section{\normalsize{SYSTEMATIC UNCERTAINTIES}}
\end{center}

Systematic effects which can lead to uncertainty in the final result
can be classified into those which affect the background estimate and
those which affect the number of expected events.  Many of the
systematic uncertainties affecting the number of expected events
depend on the top quark mass. Table II lists the different estimated
systematic uncertainties.  For the cases where there exists a top
quark mass dependence, the extreme values appear in the table.

Various effects can bias the background estimation, such as the
binning of the $\et$ distribution and the normalization method.
Dividing the $\et$ distribution into smaller bins and following the
same normalization method leads to a negligibly small difference in
the expected number of hadronic background events.  The normalization
method is based on jet trigger events, but one can check for a trigger
bias by removing the jet which was responsible for the trigger, and no
significant effect appears.  The total hadronic background is
conservatively estimated to be 17.4$\pm$2.5(stat)$\pm$0.6(sys) events,
based on these cross checks.

\begin{figure}[t]
  \parbox{3.25in}{\epsfxsize=3.25in\epsffile[60 0 320 240]{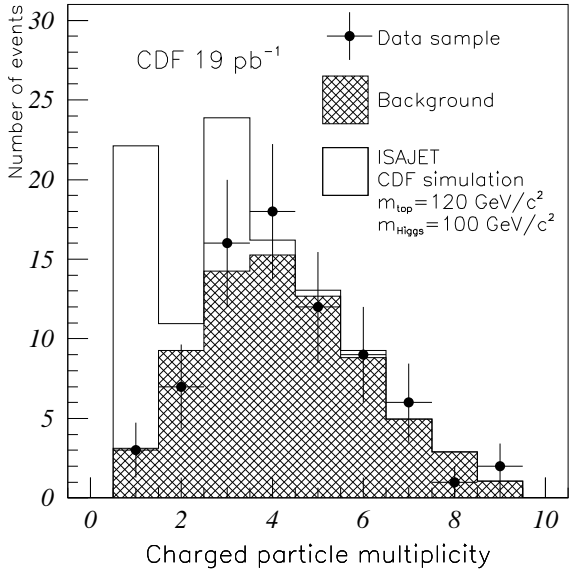}}
  \caption{Charged particle multiplicity
               distribution in the $10^{\circ}$ cone after all the cuts, for 
               the data sample (points),for the background normalized
               to the data (cross-hatched histogram), and for the expected
               signal from the normalized Monte Carlo simulations
               (open histogram) added to the  background.}
  \label{finalresult}
\end{figure}

We have compared the number of expected events for different masses of
the top quark and the charged Higgs with and without initial-state
gluon radiation in ISAJET. Half the difference between these numbers
was taken as the systematic uncertainty.  The mean value between the
number of expected events with and without initial-state gluon
radiation was taken as the number of expected events.  The isolation
cut is the criterion most affected by initial-state gluon radiation;
the number of jets is smaller with no initial-state gluon radiation.
The probability to have associated charged particles of the tau
candidate mixed with a particle of another jet becomes smaller and the
efficiency of the cut increases.  This effect also depends on the top
quark mass, since for a heavy top quark there is less energy to
produce jets as a result of the initial-state gluon radiation.

The systematic uncertainty on the trigger efficiency was estimated by
varying each point of the measured trigger efficiency by its
uncertainty.  The relative uncertainty on the number of expected
events due to the systematic uncertainty in the trigger efficiency is
conservatively estimated to be 5.5\%.  In this calculation, $m_{\rm
top} = 120$ GeV/$c^2$ and $m_{\rm Higgs} = 100$ GeV/$c^2$.

The absolute energy scale uncertainty varies from $\pm$10\% at 8 GeV
to $\pm$3\% at 100 GeV. In the Monte Carlo simulations, we shifted the
jet energy scale by these values, and repeated the analysis,
reconstructing the $\et$ of each jet and other relevant event
parameters. We used the mean relative difference of the change in the
number of expected events when the energy scale is shifted as the
systematic uncertainty on the energy scale.

\begin{table}[t]
  \begin{center}
  \vspace{0.2cm}
  \begin{tabular}{lc}                                    \hline\hline
    Uncertainty               &     Value      \\ \hline
    Top quark cross section   &     30--10\%   \\ 
    ISAJET gluon radiation    &     16--1.3\%  \\ 
    Integrated luminosity     &      3.7\%     \\ 
    Trigger efficiency        &      5.5\%     \\ 
    MC statistics             &    10--4\%     \\
    Energy scale effect       &    32--7.5\%   \\  
    Background estimation     &      14\%      \\ \hline \hline 
  \end{tabular}                                       
  \vspace{0.2cm}
  \caption{Sources and magnitudes of the systematic uncertainties in
           the analysis.  The values are the relative uncertainties 
           in the number of expected events, and represent the extremes 
           for the top mass range 100-174 GeV/$c^2$.}
  \end{center}
  \label{Sys_err}
\end{table}

\typeout{Start results section}

\begin{center}
\section{\normalsize{RESULTS}}
\end{center}

After selection, there remain 74 events from the data sample, of which
a total of 19 events have a tau candidate with either 1 or 3
associated charged particles.  Fig.~\ref{finalresult} shows the
multiplicity distribution for the data sample and the hadronic
background normalized to the data.  For comparison the plot shows the
distribution from the Monte Carlo signal simulation normalized to the
total integrated luminosity and added to the hadronic background
estimation.  The estimated total number of background events is
18.5$\pm$2.6, where the error comes from adding in quadrature the
systematic and statistical uncertainties.

The mass limits must take into account the uncertainties, both
statistical and systematic, on the number of expected background and
signal events.  For a given mass point, a simple Monte Carlo generates
a large ensemble of trials with the numbers of expected signal and
background events varying in a Gaussian fashion about the mean.  In
each trial it generated a number of observed events from a Poisson
distribution with a mean equal to the number of signal plus
background.  The standard deviations of the Gaussians are the combined
statistical and systematic uncertainties.  The mass point can be
excluded with 95\% confidence if in 95\% or more of the trials the
total number of events exceeds the 19 events actually observed.

\begin{figure}[t]
  \parbox{3.25in}{\epsfxsize=3.25in\epsffile[60 0 320 240]{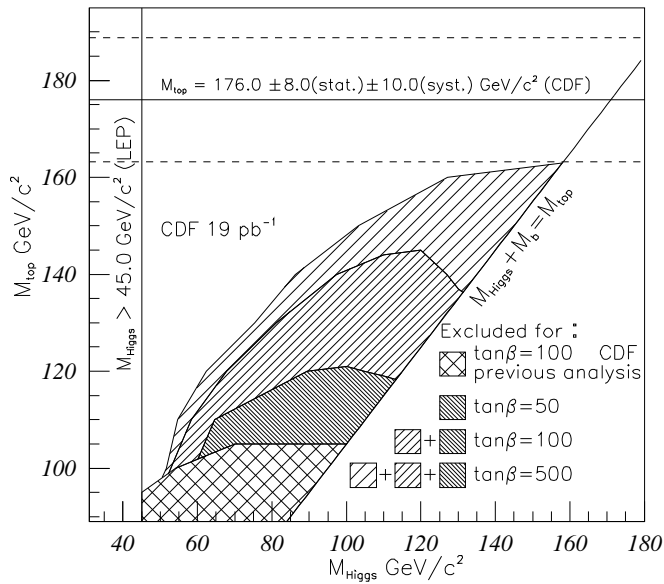}}
  \caption{Regions of the ($m_{\rm top}$,$m_{\rm Higgs}$) plane
           excluded at 95\% CL for different values of $\tan\beta$. The
           plot also shows the limit from the previous 
           analyses~{\protect \cite{CDF2164,CDF2435}}.}
  \label{masslow}              
\end{figure}

Fig.~\ref{masslarge} shows the resulting limit for large values of the
parameter $\tan\beta$, for which the branching ratios of $t\rightarrow
Hb$ and $H\rightarrow\tau\nu$ approach unity.  Fig.~\ref{masslow}
shows the limit for $\tan\beta =$ 50, 100, and 500 in the plane of the
top quark mass versus the charged Higgs mass.  As the charged Higgs
mass decreases, the missing transverse energy decreases on average,
reducing the efficiency of the selection.  Also, as the top mass
increases, its production cross section decreases, and the number of
expected events decreases. As the parameter $\tan\beta$ increases, the
branching ratios of the top to charged Higgs and charged Higgs to tau
both increase, allowing a better limit.

The event selection used is well optimized for large masses of the top
quark and the charged Higgs.  The present statistics exclude the
region for large $\tan\beta$, extending from the limit of the previous
analyses just up to the region where the top mass has been measured.

\begin{center}
\section{\normalsize{ACKNOWLEDGEMENTS}}
\end{center}

We thank the Fermilab staff and the technical staffs of the
participating institutions for their vital contributions.  This work
was supported by the U.S. Department of Energy, the National Science
Foundation, the Istituto Nazionale di Fisica Nucleare of Italy, the
Ministry of Education, Science and Culture of Japan, the Natural
Sciences and Engineering Research Council of Canada, the National
Science Council of the Republic of China, and the A. P. Sloan
Foundation, and the Alexander von Humboldt-Stiftung.


\begin{thebibliography} {99}
\bibitem {define_eta}

The pseudorapidity $\eta$ is defined as
$\eta\equiv-\ln(\tan\frac{\theta}{2})$ with the angle $\theta$ between
the beam axis and the direction of a particle measured assuming a
z-vertex position of zero unless otherwise noted, where z is the
position along the beam axis.
\end{thebibliography}
\end{document}